\documentclass{elsart1p}

\usepackage[pdftex]{graphicx}
\usepackage{units}
\usepackage{slashed}
\usepackage{bm,bbm}
\usepackage{amsmath,amssymb,amsfonts}
\usepackage{mathrsfs}
\usepackage{ucs}
\usepackage[utf8x]{inputenc}
\usepackage{hyperref,url}

\def\msun{\ensuremath{M_{\odot}}}

\def\msun{\ensuremath{M_{\odot}}}

\begin{document}
\begin{frontmatter}
 \title{Quark matter influence on observational properties of compact stars}
 \author{Sjoerd Hardeman}
 \address{Instituut-Lorentz for Theoretical Physics, Niels Bohrweg 2, NL-2333 CA Leiden, The Netherlands}
 \begin{abstract}
 Densities in compact stars may be such that quarks are no longer confined in hadrons, but instead behave as weakly interacting particles. In this regime perturbative calculations are possible. Yet, due to high pressures and an attractive channel in the strong force, condensation of quarks in a superfluid state is likely. This can have interesting consequences for magnetic fields, especially in relation to the discovery of slow-period free precession in a compact star. In this proceedings there will be a discussion of the mass-radius relations of compact stars made from quark matter and magnetic field behaviour in compact stars with a quark matter core.
\end{abstract}

\begin{keyword}
Dense quark matter \sep Quark star observational properties
\PACS 11.10.Wx \sep 12.38.Bx \sep 21.65.Qr \sep 26.60.Kp
\end{keyword}
\end{frontmatter}

\maketitle

\section{Introduction}
In the collapse of a star more massive than about $8\msun$, it is likely that the remaining compact object will be more massive than the Chandrasekhar limit, and will collapse to a compact star with a very high central density or to a black hole. A compact star created from such an event is characterised by a mass $M \sim \msun$ and a radius $R \sim \unit[15]{km}$, and was first proposed in 1933 by Baade and Zwicky, just after the discovery of the neutron. In such an object a neutron-proton superfluid-superconductor was expected. Hence, such objects are often named neutron stars.  However, it is possible that more exotic forms of matter, such as quark matter, are (also) present inside these objects.  To avoid confusion, the name neutron star will be reserved for an object truly made of neutrons. The term compact star will be used for any small, dense object that is not a black hole.

Due to Pauli blocking, only particles near the Fermi surface are expected to interact. At very high densities, the Fermi energy can become large enough to have a weakly coupled strong force, thus $\alpha_{s} \ll 1$. In 1976 Freedman and McLerran calculated a perturbative expansion of QCD with three massless quarks at large chemical potential to second order in the coupling $\alpha_{s}^{2}$ \cite{prd16_1169}, and argued that there should be a phase transition at high densities where quarks instead of hadrons are the degrees of freedom. As the $\bar{3}$-channel of the colour interactions is attractive, it allows for a colour superconducting state \cite{Collins:1974ky}.

The behaviour of quark matter at asymptotically high densities is quite well understood. The challenge is to understand the behaviour of matter at densities expected in compact stars, around or just above the QCD deconfinement phase transition. In principle, calculations in this regime can be directly related to observations of compact stars, and might help to improve the understanding of dense matter.

This proceedings will mainly focus on the current status in confronting calculations with observations. Section 2 discusses the mass-radius relation of quark stars, section 3 focuses on the magnetic properties of compact stars with a quark matter core.

\section{Mass-radius relation}
\begin{figure}[!htb]
 \begin{center}
  \includegraphics[width=0.5\textwidth]{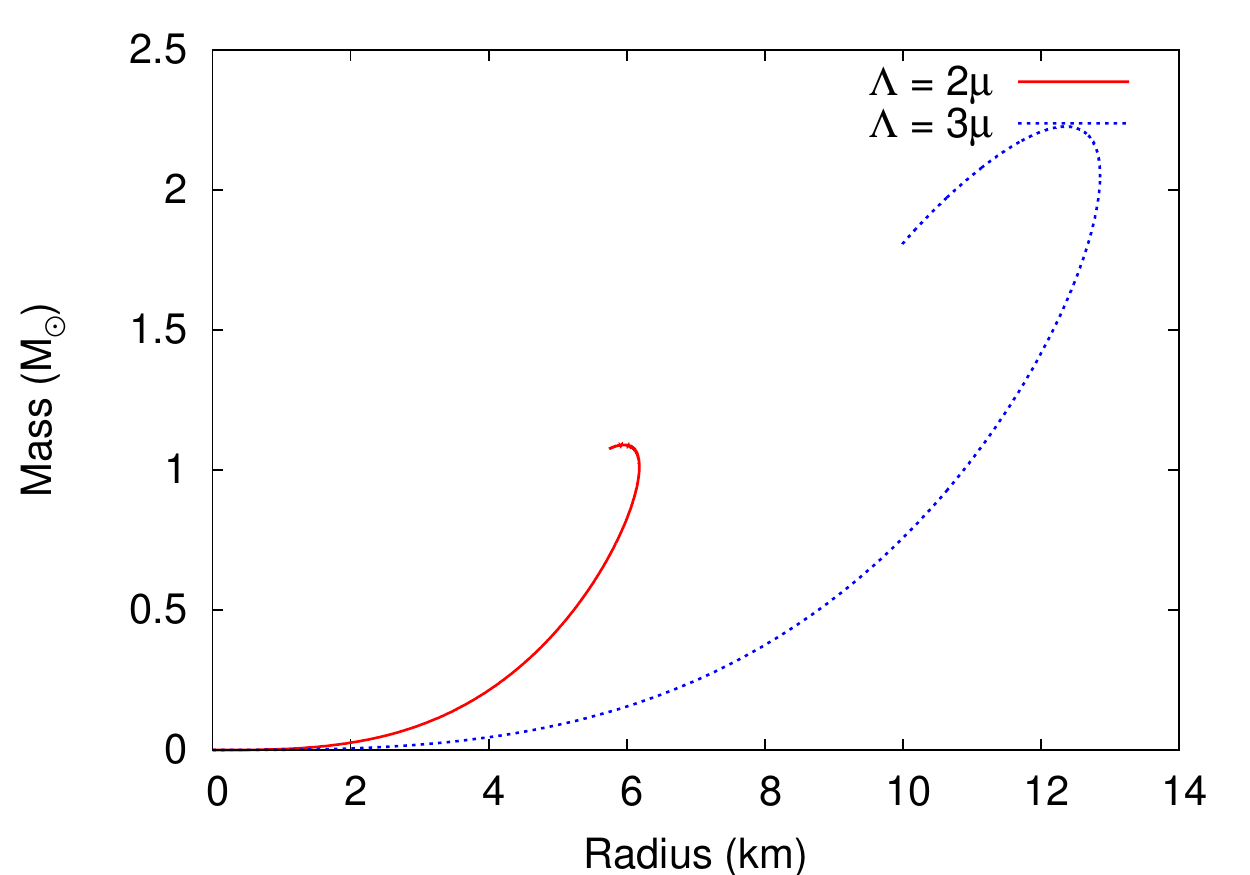}
 \end{center}
 \caption{Mass-radius relation using for a pure quark star made of massless quarks, with EoS eq. (\ref{mr}), for different values of the cutoff scale $\Lambda$.}\label{fig_2ndorder}
\end{figure}
Given an equation of state, the TOV-equations \cite{pr55_364,pr55_374} can be integrated numerically to obtain a mass-radius relation. Although the measurement of radii is observationally difficult, masses can be measured accurately in compact binaries \cite{Lattimer:2006xb,Stairs:2006yr}, suggesting an upper mass limit not much larger than $2\msun$. Interestingly, most neutron stars are found to have masses very similar to the Chandrasekhar mass of about $1.4\msun$.

As equation of state (EoS), one can use for example the two-loop perturbative QCD EoS of Freedman and McLerran. The thermodynamic potential $\Omega(\mu)$ is then given by
\begin{equation}\label{mr}
 \begin{split}
  \Omega(\mu) = -\frac{N_{f}\mu^{4}}{4\pi^{2}}&\left\{ 1-2\left(\frac{\alpha_{s}}{\pi}\right) - G + N_{f} \log \left(\frac{\alpha_{s}}{\pi}\right)\right. \\
  & + \left.\left(11-\frac{2}{3}N_{f}\right)\log \left(\frac{\bar{\Lambda}}{\mu}\right)\left(\frac{\alpha_{s}}{\pi}\right)^{2}\right\},
 \end{split}
\end{equation}
with $\alpha_{s}(\Lambda/\Lambda_{\text{QCD}})$. From this, a mass-radius relation can be calculated (figure \ref{fig_2ndorder}). The relation shown is similar to the relation found by Fraga et al. \cite{prd63_121702}. However, some differences occur in the energy density relation $\epsilon(\mu)$, see \cite{eigen} for a detailed discussion.  For nonzero strange quark mass the thermodynamic potential has been calculated to one loop order \cite{prd17_2092}. The use of a realistic strange quark mass was found to be important at one-loop level \cite{prd71_105014,eigen}. From the large scale-dependence in figure \ref{fig_2ndorder} it is clear that the convergence of the two-loop perturbative expansion is poor at the densities considered in compact stars. Perhaps a higher order expansion \cite{vuorinen08} will improve this. 

\section{Magnetism and compact star rotation}
\begin{figure}[!htb]
  \begin{center}
   \includegraphics[width=0.5\textwidth]{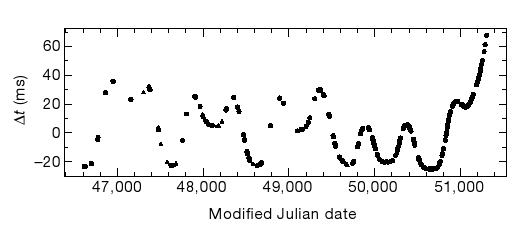}
  \end{center}
  \caption{Timing residuals from timing measurements of PSR B1828-11, with $\Delta t$ the difference of the arrival time of a pulse compared to 
  the long term mean value. The non-sinusoidal shape suggests a triaxial body \cite{2007Ap&SS.308..435L}. This image is from Stairs et al. \cite{nat406_484}, where also the Fourier transform of this plot can be found which clearly shows a peak at 500 days.\label{timing_noise}}
\end{figure}
Most compact stars are thought to start their lives as a pulsar, a rapidly rotating compact star with a very strong magnetic field that emits pulsed EM-radiation as a consequence of this. In a pulsar, the rotation and perturbations of the rotation are easily observable. It has been found that the rotation is very stable, with a predictable spindown due to magnetic braking. On top of this, perturbations such as a sudden spin-up (glitches) occur. Also, in a few sources periodic perturbations have been found, which are thought to signal slow free precession, much slower than the spin period of $\sim 1$s. The most clear cut case is PSR B1828, which has a very clear periodic signal with a period of about 500 days (figure \ref{timing_noise}). The processes that perturb the rotation are very sensitive to the internal structure of a compact star, and provide a useful tool to study it.

To be more precise, when present inside a compact star, quark matter is likely to be in one of the colour superfluid states (see \cite{Alford:2007xm} for a recent review). A significant difference with a neutron-proton superfluid/superconductor is that the probably more common colour superconducting states leave a linear combination of the photon and a gluon massless, and are therefore not electromagnetic superconductors \cite{nuphb571_269}. The difference is thus that neutron-proton matter is a superconductor, while quark matter is not.

When matter in a compact star is both a superfluid and a type II superconductor, a compact star will have fluxoids and vortices in its core. According to Link \cite{prl91_101101,aap458_881}, the fluxoids will be strongly pinned to the vortices, thereby coupling the superconductor to the superfluid. The magnetic field will be linked to both the superconducting component and the iron crust, thereby linking the rotation of the superconducting component to the crust, which is observable. The pinning of vortices to fluxoids then also links the rotation of the superfluid component to the crust. As discussed by Link this has important observational consequences, because pinned vortices strongly damp precession. As a result, slow period precession is effectively ruled out. Observations of slow period precession, as in PSR B1828, then rule out the coexistence of a type II superconductor with a superfluid in a compact star. 

Possible solutions include assuming the proton superconductor to be of type I, or to assume that the proton superconductor and neutron superfluid do not coexist. As it is problematic to assume the latter at central densities of compact stars and type I superconductivity is not very likely \cite{prc72_055801}, this points to exotic matter such as a not electromagnetic superconducting quark matter to be present in the compact star core. 

In a recent short review focussing on the classical compact star picture \cite{2007Ap&SS.308..435L}, there is a detailed discussion of precession in nuclear matter objects. It also includes a discussion of the observational evidence of neutron star precession.

\section{Conclusions and outlook}
Compact stars containing quark matter do fall into the parameter range found in observations. Both theory and observations of the most direct measure, the mass-radius relation, are problematic. However, studying the magnetic field behaviour by looking at higher orders of rotation appears to be a promising route to test the behaviour of dense matter. The observation of precession discussed here is just one possibility to test the behaviour of dense quark matter in compact stars. Perturbations such as glitches and compact star quakes are also likely to depend strongly on the interior on the compact star, and thus might shed light on the question of what is inside.

\section{Acknowledgments}
I wish to thank Daniël Boer (Vrije Universiteit) and Rudy Wijnands (Universiteit van Amsterdam) for supervision of this Master's project. Furthermore, I would like to thank Ana Achúcarro and Maarten van Hoven for useful discussions and comments. Finally, I thank the organisers for an inspiring conference. Work supported by the Netherlands Organisation for Scientific Research (NWO) under the VIDI and VICI programmes.

\bibliographystyle{h-elsevier}
\bibliography{literatuur}
\end{document}